\documentclass[twocol]{ametsocV5}

\usepackage{amsmath,amsfonts,amssymb,bm}
\usepackage{mathptmx}
\usepackage{newtxtext}
\usepackage{newtxmath}
\usepackage{amsmath}
\usepackage{grffile}

\bibpunct{(}{)}{;}{a}{}{,}

\title{Revisiting the Quasi Biennial Oscillation as Seen in ERA5.\\
Part II: Evaluation of Waves and Wave Forcing}

\authors{Hamid A. Pahlavan\correspondingauthor{Hamid A. Pahlavan, pahlavan@uw.edu}, John M. Wallace, Qiang Fu}

\affiliation{Department of Atmospheric Sciences, University of Washington, Seattle, Washington}

\extraauthor{George N. Kiladis}

\extraaffil{ Earth System Research Laboratory, NOAA, Boulder, Colorado}

\abstract{This paper describes stratospheric waves in ERA5 reanalysis and evaluates the contributions of different types of waves to the driving of the quasi-biennial oscillation (QBO). Because of its higher spatial resolution compared to its predecessors, ERA5 is capable of resolving a broader spectrum of waves. It is shown that the resolved waves contribute to both eastward and westward accelerations near the equator, mainly by the way of the vertical flux of zonal momentum. The eastward accelerations by the resolved waves are mainly due to Kelvin waves and small-scale gravity (SSG) waves with zonal wavelengths smaller than 2000 km, whereas the westward accelerations are forced mainly by SSG waves, with smaller contributions from inertio-gravity and mixed-Rossby-gravity waves. Extratropical Rossby waves propagate into the tropical region and impart a westward acceleration to the zonal flow. They appear to be responsible for at least some of the irregularities in the QBO cycle.}

\begin{document}

\maketitle

\section{Introduction}

The Quasi-Biennial Oscillation (QBO) is the dominant mode of interannual variability in the tropical stratosphere.  It is characterized by alternating, downward propagating westerly and easterly zonal wind regimes, with a period of about 28 months. For reviews of the QBO literature, see Baldwin et al. 2001, Anstey et al. (2020), and references therein. The QBO is driven primarily by equatorial synoptic scale and gravity waves propagating from the troposphere into the stratosphere and interacting with the stratospheric mean flow (e.g., Lindzen and Holton 1968; Holton and Lindzen 1972; Dunkerton 1997). Latent heat release in precipitating clouds is the primary source of vertically propagating stratospheric equatorial waves (e.g., Salby and Garcia 1987; Stephan and Alexander 2015) that range in periods from days to minutes and include Kelvin, mixed Rossby-gravity (MRG), inertio-gravity (IG), and small-scale gravity (SSG) waves. Extratropical Rossby waves also propagate from the winter hemisphere into the tropical region and interact with the QBO (Kawatani et al. 2010a). Determining the roles and the precise partitioning between these various waves in the forcing of the QBO is still an active area of research (Holt et al. 2020).

The number of Coupled Model Intercomparison Project (CMIP) models that are able to simulate the QBO has increased from none in CMIP3 conducted ten years ago, to 5 in CMIP5 seven years ago, to 15 in the current CMIP6 (Richter et al. 2020). Most of these models simulate the period of the QBO well but underestimate its amplitude at all levels below 20 hPa. Most of the models have robust Kelvin and MRG waves in the lower stratosphere, but the forcing varies from model to model and is generally too weak and does not extend high enough (Giorgetta et al. 2002; Lott et al. 2014; Holt et al. 2020). Alexander and Ortland (2010), Ern et al. (2014), Vincent and Alexander (2020), and others have advocated that observational studies be conducted to help quantify QBO wave forcings.

Global high-resolution modeling and observational studies suggest that in the westerly shear zones of the QBO, Kelvin waves provide about half of the eastward forcing, while the rest of forcing is provided by SSG waves. In the easterly shear zones, the SSG waves provide most of the westward forcing (Kawatani et al. 2010a; Holt et al. 2020).

Given the uncertainties in the contribution of the various equatorial waves to the driving of the QBO in the models, and the unavailability of direct measurements of wave forcings, arguably the best way to proceed is to use global datasets derived from reanalyses in which all available radiosonde observations are assimilated, along with satellite-observed temperature data from 1979 onward. The ECMWF ERA-Interim (ERA-I) reanalysis has been used extensively for this purpose (e.g., Ern et al. 2014; Kim and Chun 2015) and found to be quite reliable in the tropical lower and middle stratosphere, with a good representation of planetary waves (e.g., Ern and Preusse 2009a,b; Anstey et al. 2020). However, in Part I of this study we showed that the forcing due to SSG waves in ERA-I to be negligible, in particular in the easterly shear zones, consistent with the results of Ern et al. (2014).

\begin{figure}[t]
\centerline{\includegraphics[width=15pc]{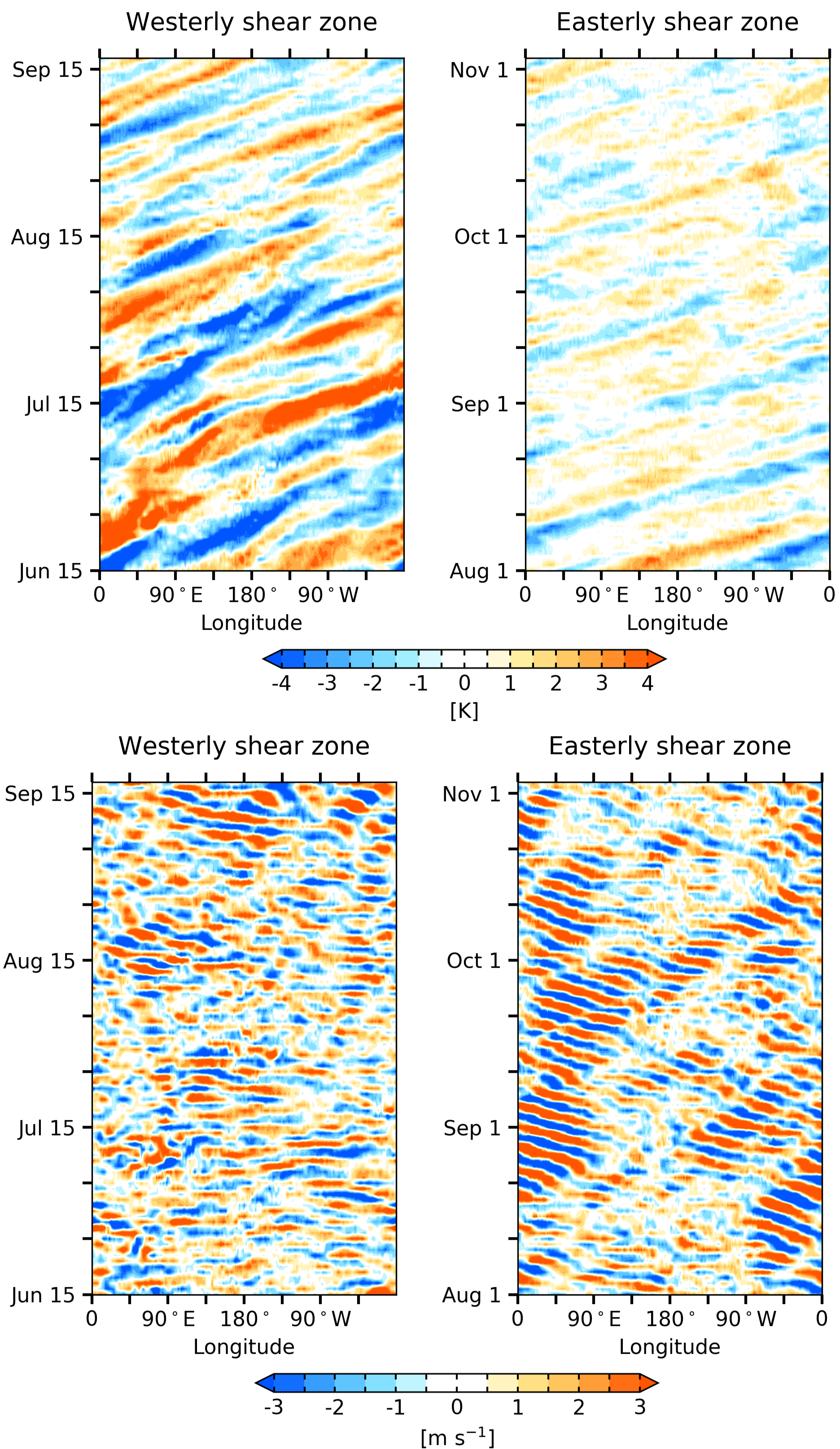}}

\caption{Equatorial time-longitude sections of daily mean temperature (top), and meridional wind (bottom), in westerly (left) and easterly (right) shear zones of the QBO at 50 hPa. The zonal mean has been removed. The year is 2015 in the left panels and 2014 in the right panels.} \label{fig1}
\end{figure}

\begin{figure}[t]
\centerline{\includegraphics[width=15pc]{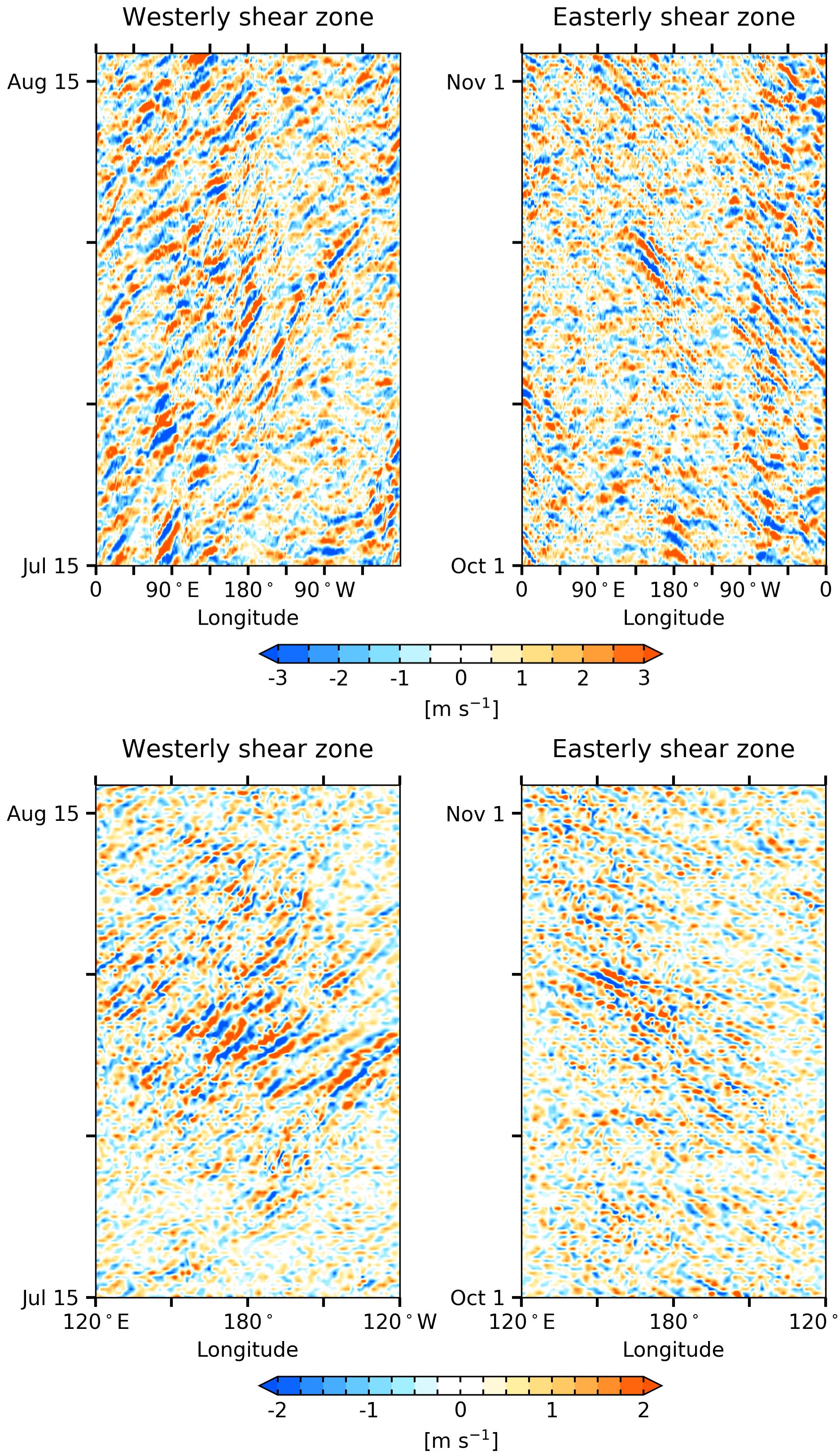}}

\caption{As in Fig. 1, but for a shorter time span of 6-hourly zonal wind over the equator. In the top panels the data have been high-pass filtered to retain frequencies higher than 0.4 cycle day$^{-1}$. In the bottom panels they are not time-filtered, but they are Fourier filtered in longitude to retain zonal wavenumbers higher than 20.} \label{fig2}
\end{figure}

Here we revisit the QBO using ERA5, the fifth generation of ECMWF atmospheric reanalyses (Hersbach et al. 2020). The higher spatial resolution in ERA5 compared to ERA-I, and in particular the higher vertical resolution makes it possible to resolve a broader spectrum of atmospheric waves and allows for a more realistic representation of wave-mean flow interaction, all of which are of crucial importance for QBO studies. For a more detailed description of ERA5 and its representation of the QBO as compared with ERA-I, we refer the reader to the Part I of this study.

In Part I, we investigated the dynamics and momentum budget of the QBO. It was shown that in ERA5, most of the QBO wave forcing is provided by resolved waves during the descent of both westerly and easterly regimes. Here in Part II, we focus on the various atmospheric wave modes and their roles in driving the QBO. Consistent with previous modeling and observational studies, we find that the eastward accelerations are mainly due to Kelvin waves and SSG waves, whereas the westward accelerations are forced mainly by SSG waves, with smaller contributions from IG and MRG waves.

This paper is organized as follows. In section 2 we analyze the equatorial winds and temperatures to reveal the various wave modes and to investigate how their distribution is mediated by the background flow in association with the QBO cycle. Section 2 also documents the two-sided wavenumber-frequency spectra of the equatorial winds and temperature in ERA5. Section 3 documents the contributions of the momentum and heat fluxes to the forcing of the QBO, as estimated by decomposing the EP flux divergence. In section 4, by making use of the two-sided spectra shown in section 2, the resolved waves are classified into Kelvin, inertio-gravity (IG), mixed Rossby-gravity (MRG), small-scale gravity (SSG, here defined as waves with zonal wavenumbers larger than 20), and extratropical Rossby waves. Then the role of each wave mode in driving the QBO is evaluated. The paper concludes with a summary and a brief discussion in section 5.

\begin{figure*}[t]
\centerline{\includegraphics[width=25pc]{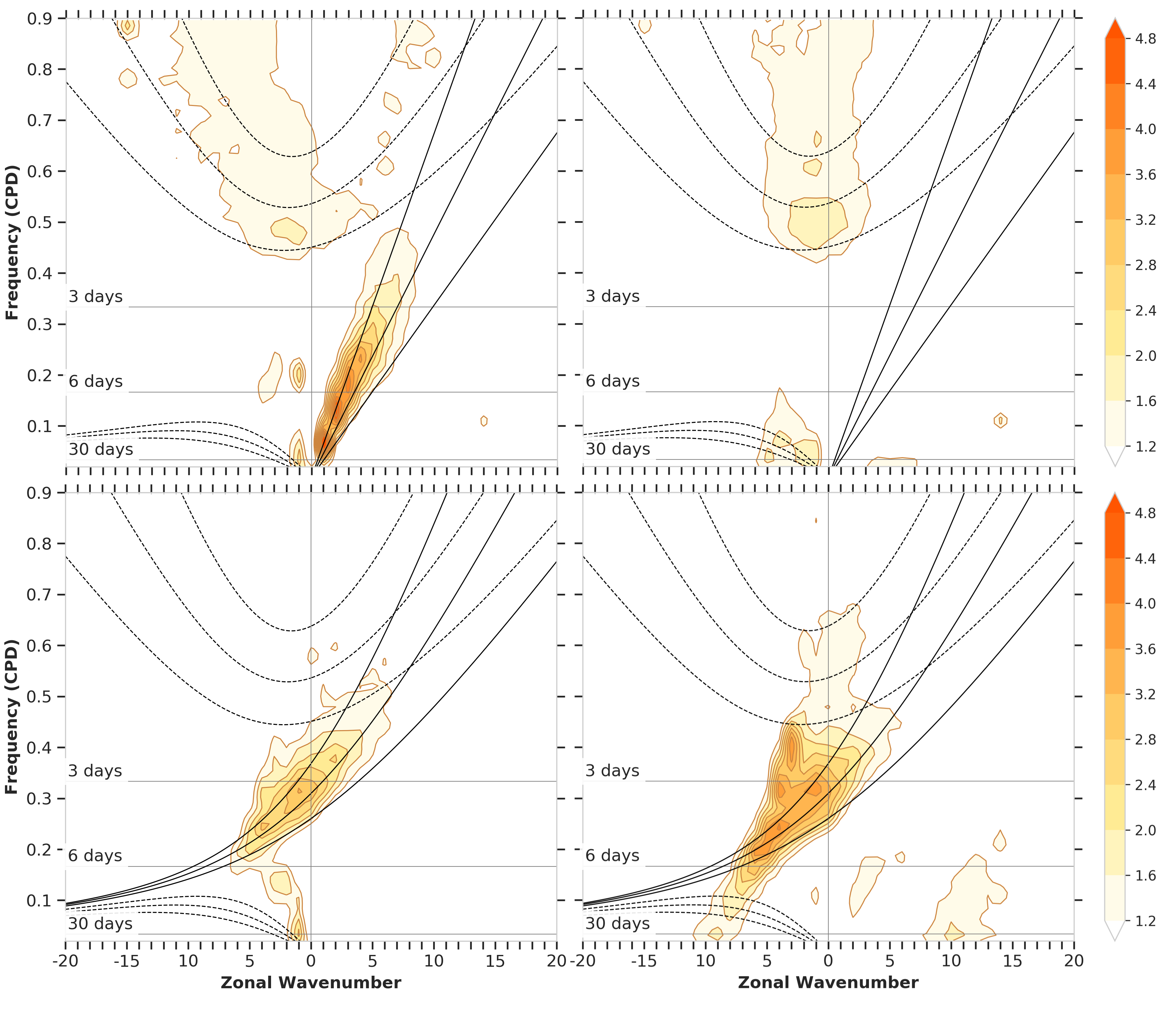}}

\caption{Power spectral density of the symmetric (top) and antisymmetric (bottom) modes of the zonal (left) and meridional (right) wind perturbations at the 50 hPa level, averaged over 15\textdegree N/S, plotted as the ratio of the computed spectrum to the background spectrum and indicated by shading. The theoretical dispersion curves for equatorial wave modes with equivalent depths of 25, 50, and 100 are superimposed.} \label{fig3}
\end{figure*}

\section{Identification of equatorial wave modes}

\subsection{Time-longitude sections}

Figure 1 (top panels) shows time-longitude sections of daily mean temperature ($T'$), where prime denotes deviation from the zonal mean, over the equator at the 50 hPa (20 km) level in a westerly shear zone (left) and in an easterly shear zone (right). Both sections are dominated by fluctuations that tilt upward toward the right, indicative of eastward propagating Kelvin waves with periods of 15 days, zonal wavenumber $k = 1, 2$, and a phase speed around $+30$ m s$^{-1}$. As expected on the basis of theory (Andrews et al. 1987), the waves in the westerly shear zone are much stronger. The bottom panels show analogous sections for meridional wind ($v'$). The waves in the westerly shear zone (left) at times appear to be standing oscillations, and overall do not have a well-defined direction of phase propagation, whereas those in the easterly shear zone (right) are clearly westward propagating disturbances with periods on the order of 4-5 days, zonal wavenumber $k\sim4$, and a phase speed of $\sim -20$ to $\sim -30$ m s$^{-1}$, consistent with MRG waves (Andrews et al. 1987; Kiladis et al. 2016). They tend to be concentrated within eastward propagating wave packets, indicative of an eastward group velocity, which is also characteristic of MRG waves.

Figure 1 is based on unfiltered daily mean data, and so is dominated by low frequency, planetary scale waves. To be able to see the waves with higher frequency and/or smaller zonal wavelength, the data need to be temporally and/or spatially filtered. The top panels of Fig. 2 show sections based on 6-hourly data that have been high-pass filtered to retain frequencies higher than 0.4 cycle day$^{-1}$: zonal wind ($u’$) over the equator at the 50 hPa level in a westerly shear zone (left) and in an easterly shear zone (right). Although they are noisy, these sections are dominated by eastward and westward propagating IG waves, respectively. Filtering 6-hourly $u’$ to retain zonal wavenumbers $|k| > 20$ reveals a prevalence of eastward propagating SSG waves in the westerly shear zones and westward propagating waves in the easterly shear zones, respectively. Filtering the data to separate eastward and westward propagating IG and SSG waves yields a similar result (see Fig. S1).

\begin{figure*}[t]
\centerline{\includegraphics[width=35pc]{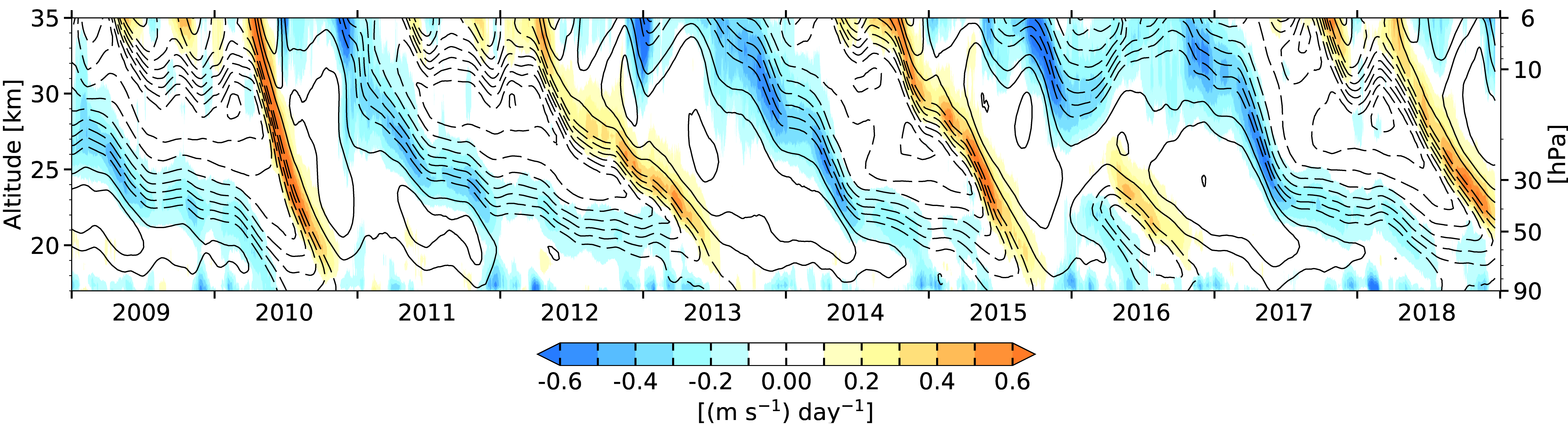}}

\caption{Time-height section of 6-hourly EP flux divergence (colored shading), and zonally averaged zonal wind (contours), averaged within 5\textdegree  of the equator. Contour interval 5 m s$^{-1}$, westerlies solid, easterlies dashed, zero contour omitted. Based on 6-hourly data, smoothed by a 30-day running mean.} \label{fig4}
\end{figure*}

\begin{figure}[t]
\centerline{\includegraphics[width=19pc]{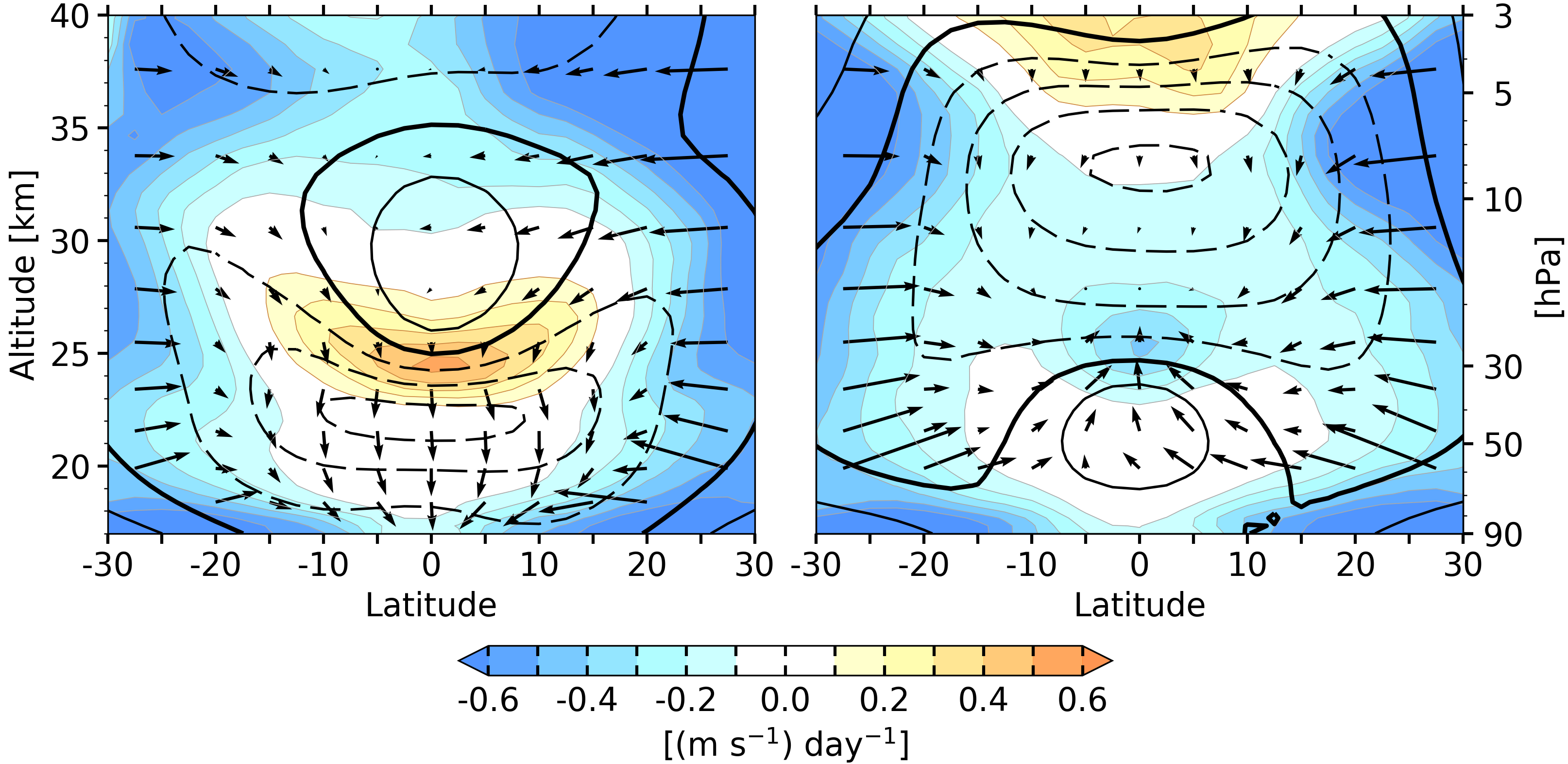}}

\caption{Composite fields of EP flux divergence (colored shading), EP flux (vectors), and zonally averaged zonal wind (contours) for descending westerly (left) and easterly (right) shear zones of the QBO. In the westerly shear zone (left), the longest vertical components of the arrows correspond to $5.2 \times 10^8$ kg s$^{-2}$, while the longest horizontal components correspond to $1.5 \times 10^{11}$ kg s$^{-2}$. In the easterly shear zone (right), the arrows are scaled to be twice as long. Contour interval 7.5 m s$^{-1}$, westerlies solid, easterlies dashed, zero contour bolded.} \label{fig5}
\end{figure}

\subsection{Wavenumber-frequency spectrum}

Figure 3 shows the power spectra of $u'$ and $v'$ at 50 hPa, averaged over 15\textdegree  N/S, plotted as a function of zonal wavenumber and frequency after removing the background spectrum following Wheeler and Kiladis (1999). The variables are split into symmetric and antisymmetric components with respect to the equator. The analysis is based on 40 years (1979-2018) of 6-hourly data with a 96-day window and 65-day overlap. The theoretical dispersion curves of the equatorial wave modes (Matsuno 1966; Andrews et al. 1987) are superimposed on the spectra, for several different equivalent depths (h).

Most of the spectral power in the symmetric zonal wind spectrum appears along the Kelvin wave dispersion curves at $h\sim100$m with a corresponding phase speed of $\sim30$ m s$^{-1}$. The dominant features in the antisymmetric spectra of both the zonal and meridional wind components are associated with MRG waves. The band of enhanced power follows the dispersion curve for $n =$ 0 toward higher frequencies, well into the domain of eastward propagating IG waves. The MRG wave signal is mostly confined to $0.1 < \omega < 0.5$ cycle day$^{-1}$, and its peak has periods in the 2.5-5 day range. Enhanced power associated with equatorial westward IG waves is evident in the symmetric spectra of the zonal wind. The corresponding power spectra of temperature and vertical velocity shown in Fig. S2 are consistent with Fig. 3.

These results are generally consistent with the findings of previous observational and modeling studies. For example, Kim et al. (2019) showed that the spectrum of zonal and meridional winds at 50 hPa is very similar among several reanalysis products. 

\section{Term-by-term breakdown of the wave forcing}

\begin{figure*}[t]
\centerline{\includegraphics[width=35pc]{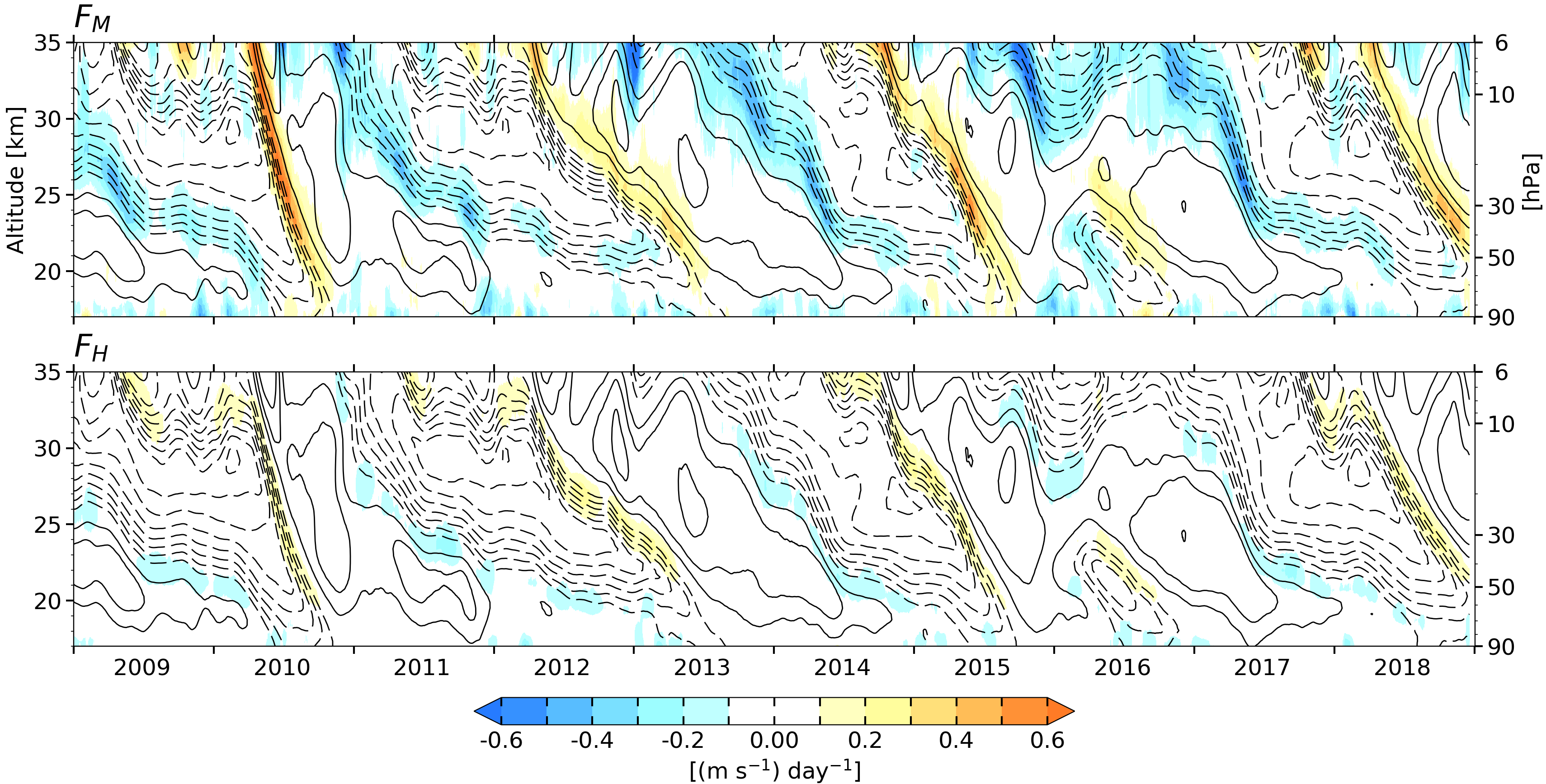}}

\caption{As in Fig. 4, but the total forcing is decomposed into the contributions of the momentum fluxes (top) and heat fluxes (bottom).} \label{fig6}
\end{figure*}

\begin{figure}[t]
\centerline{\includegraphics[width=19pc]{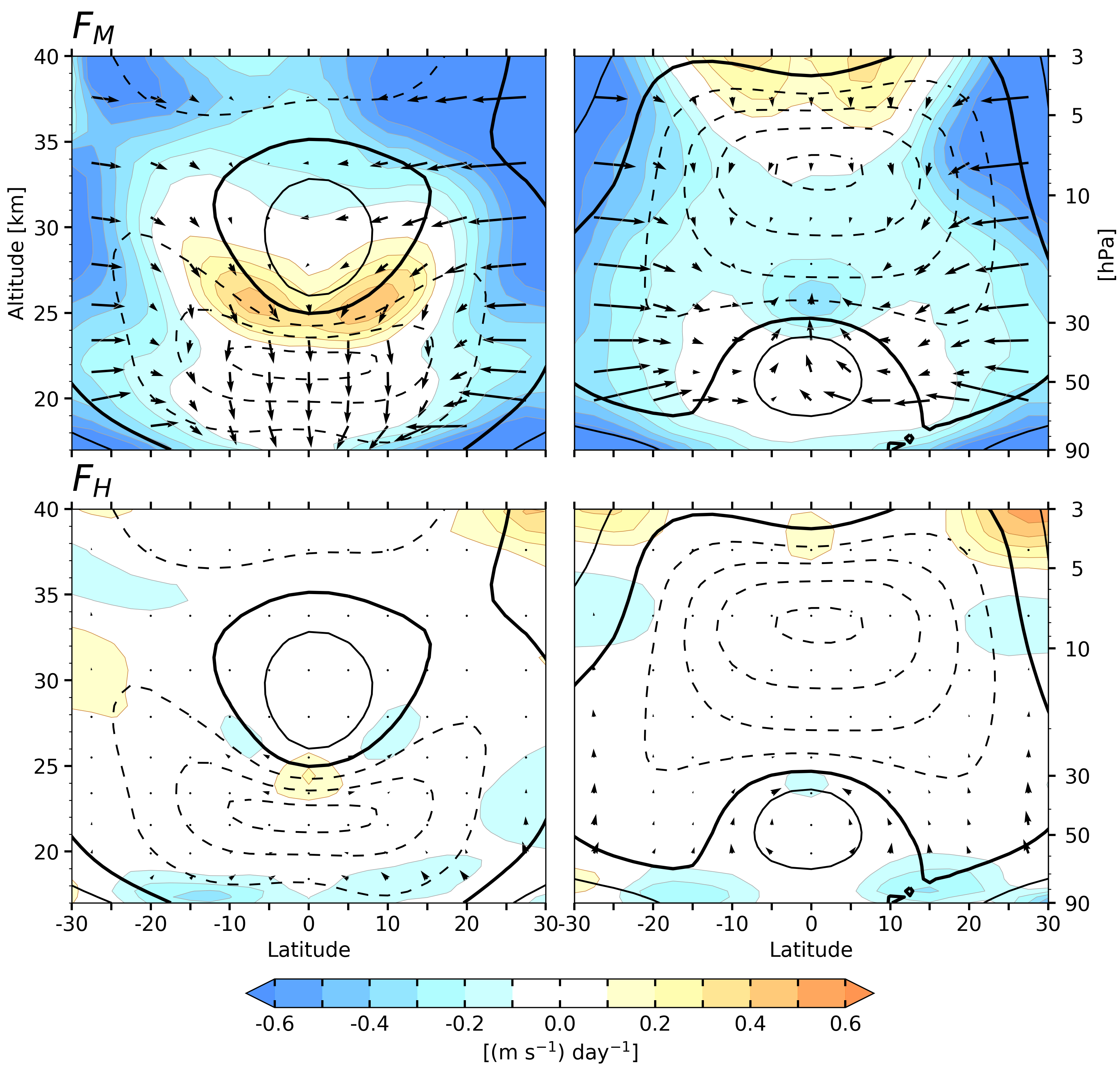}}

\caption{As in Fig. 5, but the total EP flux is decomposed into the contributions of the momentum fluxes (top) and heat fluxes (bottom).} \label{fig7}
\end{figure}

In Part I of this study we evaluate the QBO momentum budget and show the total forcing by the resolved waves based on the transformed Eulerian mean (TEM) framework (Andrews et al. 1987), in which the wave forcing of the mean flow is proportional to the divergence of the Eliassen Palm (EP) flux vector. For reference, Fig. 4 shows a time-height section of the total wave forcing near the equator during a 10-yr interval superimposed upon the distribution of zonal wind itself and Fig. 5 shows the corresponding meridional cross sections at the times of descending westerly (left) and easterly (right) shear zones in the QBO, both repeated from Part I. The composites were constructed by averaging the months in which the zero-wind line of zonally-averaged zonal wind over the equator in descending westerly or easterly shear zones reaches a prescribed reference level, here taken to be 25 hPa ($\sim$25 km). The zonal-mean forcing due to EP flux divergence is seen to be generally concentrated near the zero-wind line in the areas of highest vertical shear of the zonal winds, especially the eastward forcing in the westerly shear zone. We refer the reader to the Part I for a more detailed description of these figures.

The forcing due to EP flux divergence can be decomposed into the contributions of the momentum fluxes ($F_M= F_M^{(z)}+ F_M^{(\phi)}$), and heat fluxes ($F_H= F_H^{(z)}+ F_H^{(\phi)}$), where:

\smallskip

\hspace*{-9pt}$F_M^{(z)} =  \rho_0^{-1} \frac{\partial}{\partial z}[\rho_0(-\overline{w'u'})]  \hfill (1)$

\hspace*{-9pt}$F_M^{(\phi)} =  (a \cos^2 \phi)^{-1} \frac{\partial}{\partial \phi}[\cos^2\phi(-\overline{v'u'})]  \hfill (2)$

\hspace*{-9pt}$F_H^{(z)} =  \rho_0^{-1} \frac{\partial}{\partial z}[\rho_0(\{f - (a \cos \phi)^{-1} (\bar u \cos \phi)_\phi\} \overline{v'\theta'} / \bar\theta_z)] \hfill (3)$

\hspace*{-9pt}$F_H^{(\phi)} =  (a \cos^2 \phi)^{-1} \frac{\partial}{\partial \phi}[\cos^2\phi  (\bar u_z \overline{v'\theta'} / \bar\theta_z)]  \hfill (4)$

\smallskip

The notation follows the conventions described in Part I. We will refer to $F_M$ as the direct forcing by way of the wave fluxes of zonal momentum and to $F_H$ as the indirect forcing by way of the mean meridional circulations induced by the poleward heat transports.

\begin{figure*}[t]
\centerline{\includegraphics[width=35pc]{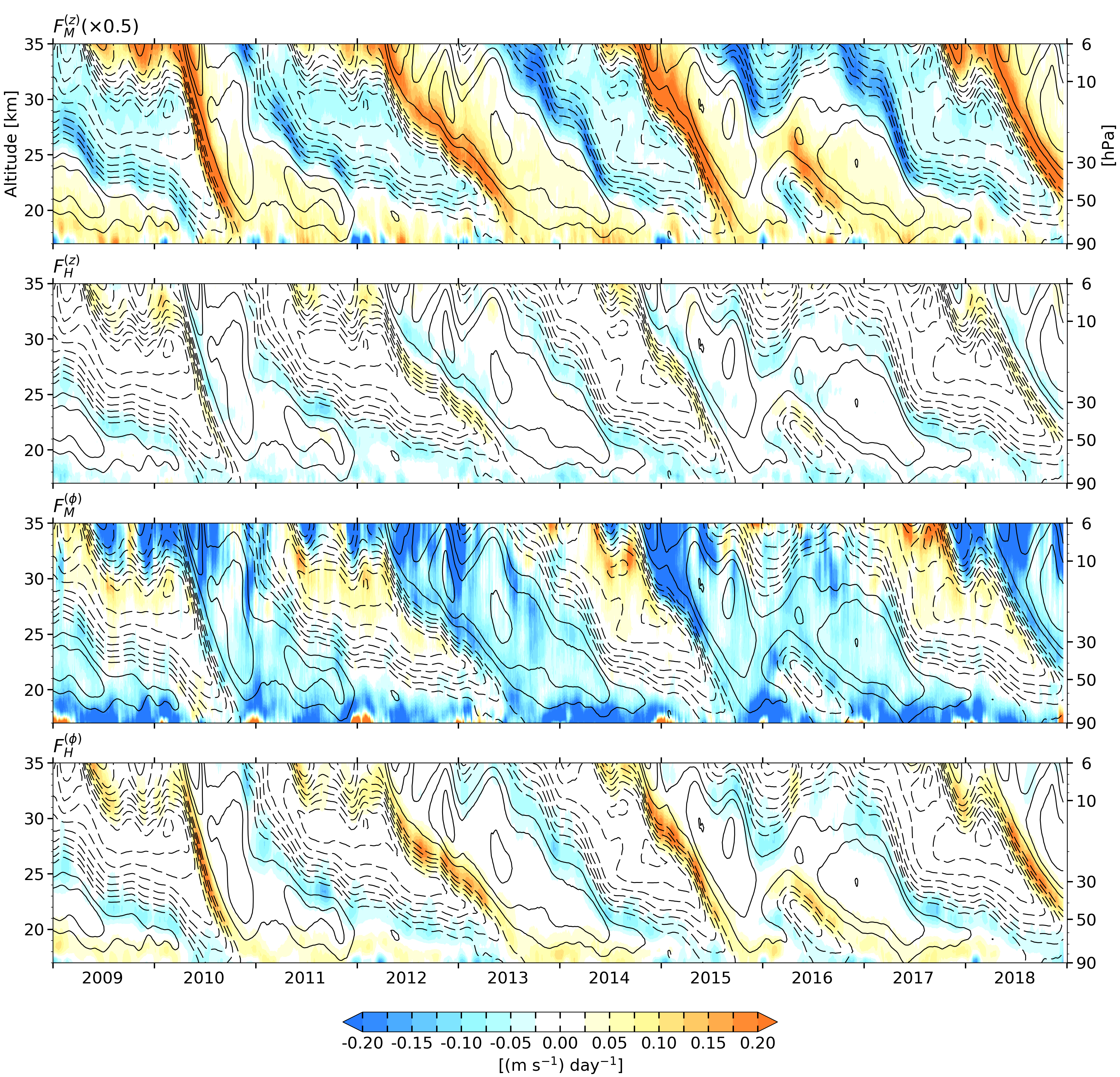}}

\caption{As in Fig. 4, but the total flux is decomposed into the contributions of each term of the EP flux in Eqns. (1-4), as indicated.} \label{fig8}
\end{figure*}

\begin{figure*}[t]
\centerline{\includegraphics[width=38pc]{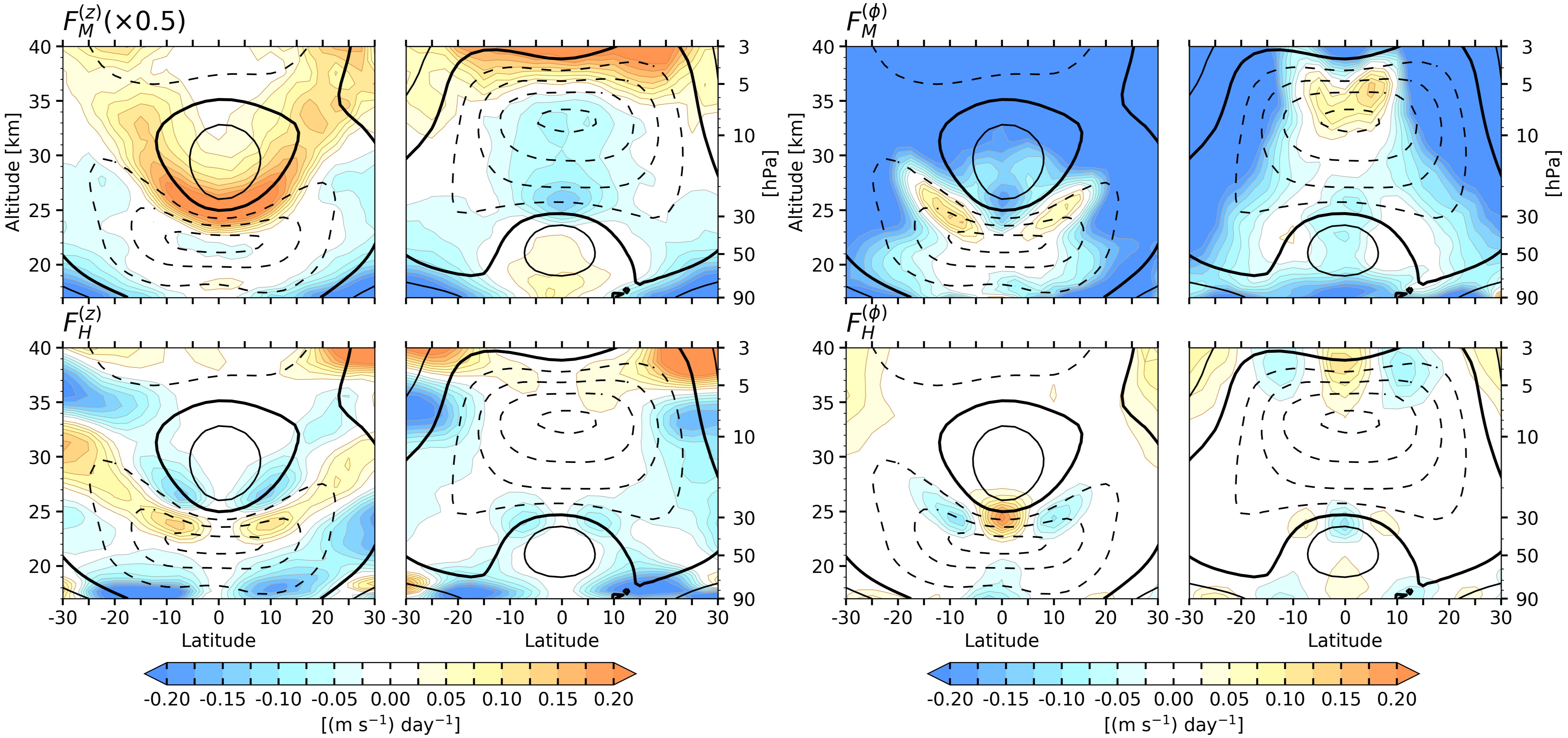}}

\caption{As in Fig. 5, but the total flux is decomposed into the contributions of each term of the EP flux in Eqns. (1-4), as indicated.} \label{fig9}
\end{figure*}

Figure 6 shows time-height sections of $F_M$ and $F_H$ and Fig. 7 shows their meridional distributions in descending westerly and easterly shear zones. Also shown in Fig. 7 is the transport of westerly momentum in the meridional plane, which is in the opposite direction of the EP flux vectors. Note that the directions of the EP flux and wave propagation in the meridional cross sections are the same for waves with westward (intrinsic) phase velocities, whereas they are in opposite directions for waves with eastward phase velocities (Andrews et al. 1983). Hence, during the descent of both westerly and easterly phases, the waves are dispersing upward near the equator, but the EP flux is downward during descending westerly phase, as they are dominated by eastward propagating waves.

It is evident that $F_M$ is the dominant contributor to the EP flux forcing and that it plays an essential role in driving the QBO. It is strongest in the shear zones, where it reaches values as high as $\pm$0.6 m s$^{-1}$ d$^{-1}$ at times of peak accelerations. The top panels of Fig. 7 show that the eastward accelerations are concentrated within descending westerly shear zones, with off-equatorial maxima. In contrast, weak westward accelerations extend throughout much of the of easterly wind regimes, maintaining them in the presence of poleward flow induced by radiative damping of the QBO-related temperature perturbations, as described in Part I. The main purpose of including $F_H$ in Figs, 6 and 7 is to demonstrate that it is only of minor importance in the forcing of the QBO. However, its structure in the vicinity of descending westerly shear zones, with eastward forcing over the equator flanked by westward accelerations, render the meridional structure of the total forcing smoother than that of the $F_M$ forcing, with a maximum over the equator, as shown in Fig. 5.

\begin{figure*}[t]
\centerline{\includegraphics[width=35pc]{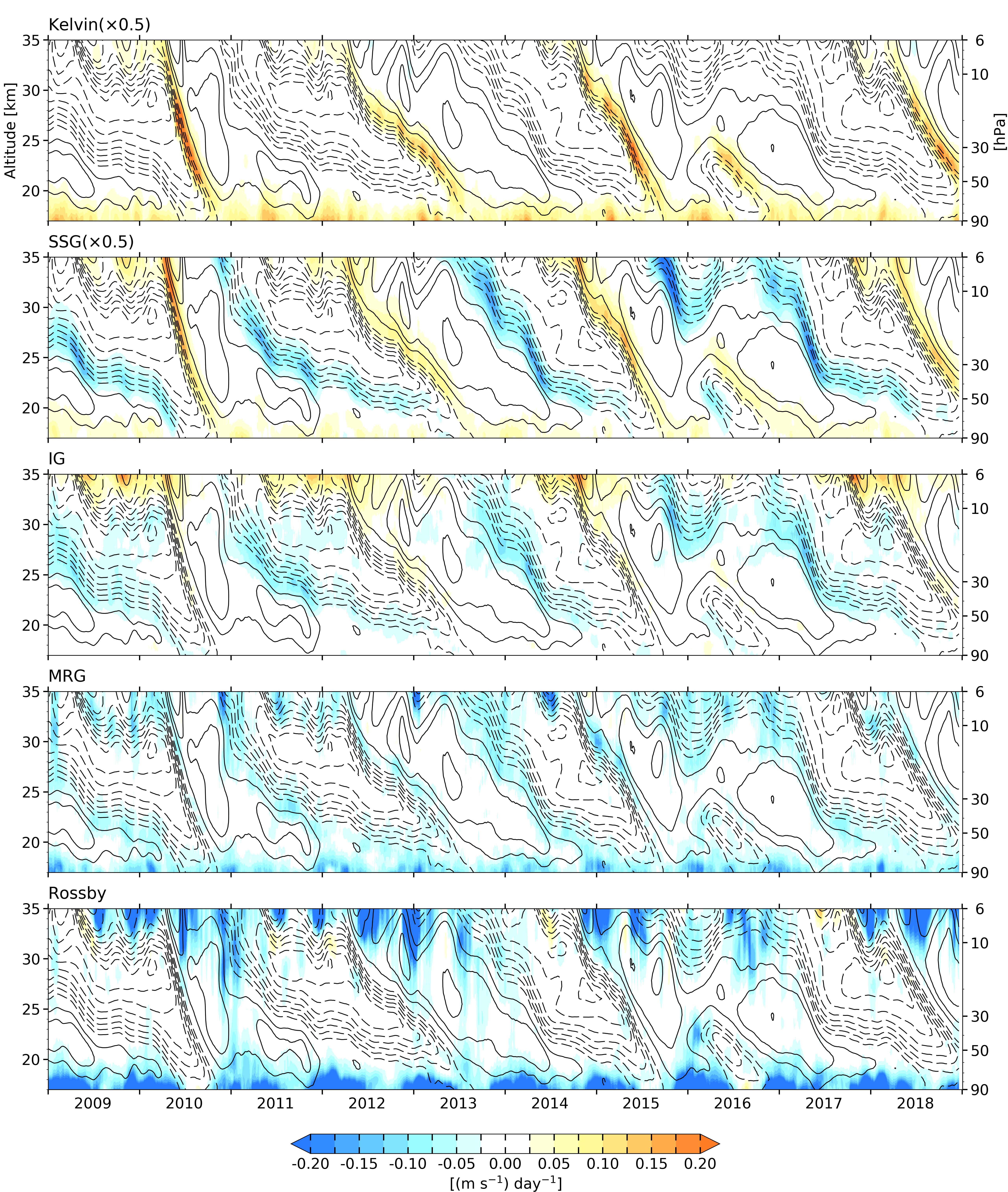}}

\caption{As in Fig. 4, but the total flux is decomposed into the contributions of each term of the EP flux in Eqns. (1-4), as indicated.} \label{fig10}
\end{figure*}

\begin{figure*}[t]
\centerline{\includegraphics[width=38pc]{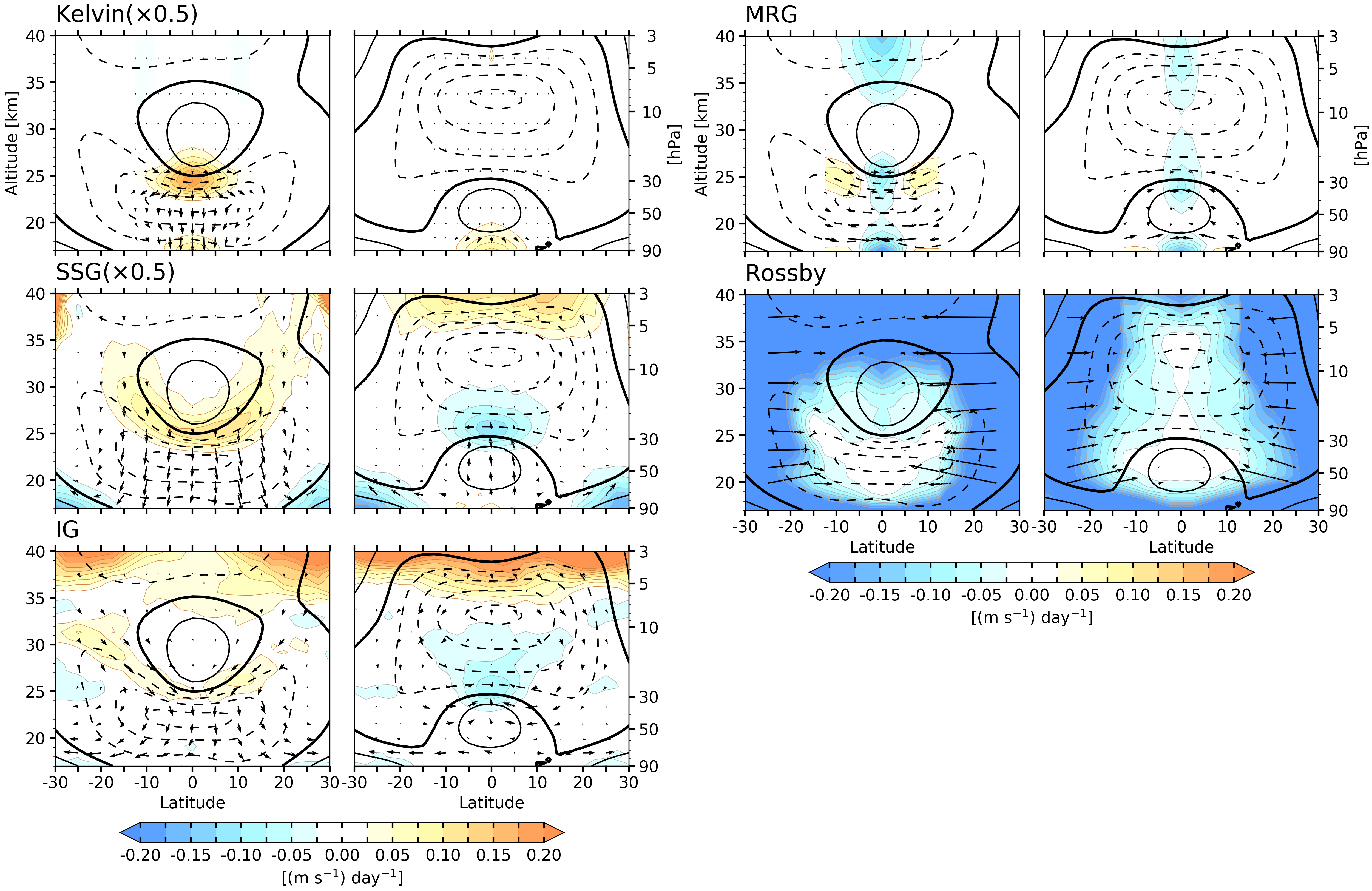}}

\caption{As in Fig. 5, but the total flux is decomposed into the contributions of each type of wave, as indicated. The longest vertical components of the arrows correspond to $9 \times 10^7$ kg s$^{-2}$, while the longest horizontal components correspond to $1.3 \times 10^{11}$ kg s$^{-2}$. For the Rossby waves, the arrows are scaled to be twice as short.} \label{fig11}
\end{figure*}

The corresponding four-term decomposition based on Eqns. (1)-(4) is shown in Figs. 8 and 9. The term $F_M^{(z)}$ involving the vertical fluxes of zonal momentum is by far the largest of the four terms (note that the contour interval is doubled relative to that of the other terms) and is also the dominant contributor to $F_M$ in Figs. 6 and 7. The heat flux term $F_H^{(z)}$, is zero on the equator, where the Coriolis force vanishes, but it exhibits highly structured off-equatorial features, with dipoles of accelerations in the vicinity of the westerly shear zone with maxima $\sim$10\textdegree  N/S, which project weakly upon the equatorial time-height section shown in Fig. 8.  It contributes to the descent of the easterly shear zones and it tends to spread the easterly wind regimes into subtropical latitudes while weakening them at lower latitudes.

$F_M^{(\phi)}$ is dominated by the westward acceleration induced by the poleward transport of westerly momentum ($v'u'$) by the extratropical waves in the winter hemisphere, which is evidently modulated by the QBO. The westward accelerations protrude deeper into the tropics in the westerly regimes, thereby weakening them. The flux divergence exhibits distinct equatorial maxima, which also act to weaken the westerly regimes. It also exhibits a rather intricate pattern within and just below the westerly shear zones consisting of a westward forcing on the equator, flanked by eastward forcing centered $\sim$10\textdegree N/S. Hence, this term acts to retard the descent of the westerlies close to the equator while accelerating it away from the equator, thereby reducing the curvature of the shear zone and broadening the westerly regime and preventing it from becoming inertially unstable. It is notable that  $F_H^{(\phi)}$ displays a similar pattern in the westerly shear zone that is of roughly comparable magnitude but opposing sign.

\section{Wave-by-wave breakdown of the forcing}

The forcing attributable to each type of wave is calculated after splitting the perturbation variables in Eqns. (1)-(4) within the domain 12.5\textdegree  N/S into equatorially symmetric and anti-symmetric components. Each component is then spectrally transformed into the zonal wavenumber-frequency $(k-\omega)$ domain and filtered to retain eastward and westward propagating waves in prescribed domains. Following conventions used in previous studies such as Kim and Chun (2015a), and based on the spectral analysis above (Figs. 3 and S2), the perturbations for the Kelvin waves are restricted to $1 \leq k \leq 20$ and $\omega < 0.4$ cycle day$^{-1}$ in the symmetric spectrum, and those for the MRG waves are restricted to $|k| \leq 20$ and $0.1 \leq \omega \leq 0.5$ cycle day$^{-1}$ in the anti-symmetric spectrum. Hence in this paper, the MRG waves refer to both the westward and eastward propagating (EIG) $n = 0$ waves, where n is the index of the Legendre polynomial, as both components comprise a continuum of wave activity as discussed in Kiladis et al. (2016), and Dias and Kiladis (2016). Spectral components that are not defined as Kelvin and MRG waves are considered as Rossby waves if $|k| \leq 20$ and $\omega < 0.4$ cycle day$^{-1}$, and IG waves if $|k| \leq 20$ and $\omega \geq 0.4$ cycle day$^{-1}$. Perturbations with $|k| > 20$ (i.e., wavelengths shorter than 2000 km) are classified as SSG waves regardless of their frequency and direction of propagation.

The equatorial time-height section and the meridional cross section for the forcing by each of the five wave modes are shown in Figs. 10 and 11, respectively and the contributions of the heat and momentum fluxes to the total forcing of each wave mode in the westerly and easterly shear zones are shown in Figs. 12 and 13, respectively.

The eastward acceleration attributable to the Kelvin wave is concentrated in the westerly shear zones (Fig. 10). It exhibits a nearly Gaussian shape in latitude centered on the equator (Fig. 11), and mainly results from the vertical flux of zonal momentum (Fig 12). The Kelvin wave forcing dominates the eastward acceleration in the middle and lower stratosphere, reaching values as high as +0.5 m s$^{-1}$ d$^{-1}$ in strong shear zones, consistent with previous estimates. For example, Alexander and Ortland (2010) used temperature measurements from the High Resolution Dynamics Limb Sounder (HIRDLS), and estimated that about half of the acceleration in the eastward shear zone is due to Kelvin waves, with a typical magnitude of about +0.5 m s$^{-1}$ d$^{-1}$. Similarly, Kim and Chun (2015b), using ERA-I model level data at 30 hPa, found that the Kelvin wave dominates the resolved forcing in the westerly shear zone, with accelerations of about 7-13 m s$^{-1}$ month$^{-1}$ ($\sim0.2-0.4$ m s$^{-1}$ d$^{-1}$).

Also shown in Fig. 11 are the EP fluxes. As it is mentioned earlier, the direction of the EP flux and wave propagation are opposed to each other for waves with eastward phase velocity such as Kelvin waves, and eastward propagating SSG and IG waves. Hence the corresponding EP fluxes are downward in the westerly shear zone composites because the waves are dispersing upward (Fig .11). Note that the easterly jet in the lower stratosphere provides a favorable environment for the upward dispersion of eastward propagating waves originating in the troposphere.

It is evident from Fig. 11 that during the descent of the westerly regimes (left), the Kelvin waves disperse upward through the easterly regime in the lower stratosphere and into the middle stratosphere. Note that Kelvin waves can disperse upward through an easterly background flow without appreciable dissipation. Kelvin waves are absorbed when they encounter a westerly shear zone, imparting a strong eastward acceleration. When the westerly jet is located in the lower stratosphere (right panel), the Kelvin waves dissipate in the lower stratosphere, and thus, their amplitudes in the middle and upper stratosphere are very small. It is also evident from Fig. 11 that as the Kelvin waves dissipate within the westerly shear zone, their EP fluxes tend to diverge outward from the equatorial belt, consistent with results from Kawatani et al. (2010) and Kim and Chun (2015b).

The SSG waves produce eastward accelerations in the westerly shear zones and westward accelerations in easterly shear zones of comparable magnitude, and appear to be primarily responsible for the episodes of rapid descent of easterly shear zones (Fig. 10) (Part I). It is evident from Fig. 11 that during the descent of the westerly shear zones, the waves propagate upward through the easterly regime over a broad range of latitudes and converge into the westerly shear zone, inducing an eastward acceleration over a broad arc in the meridional plane. During the descent of the easterly shear zones, the SSG waves propagate through the narrow westerly jet and dissipate in the easterly shear zone above it, inducing a westward acceleration.

\begin{figure*}[t]
\centerline{\includegraphics[width=35pc]{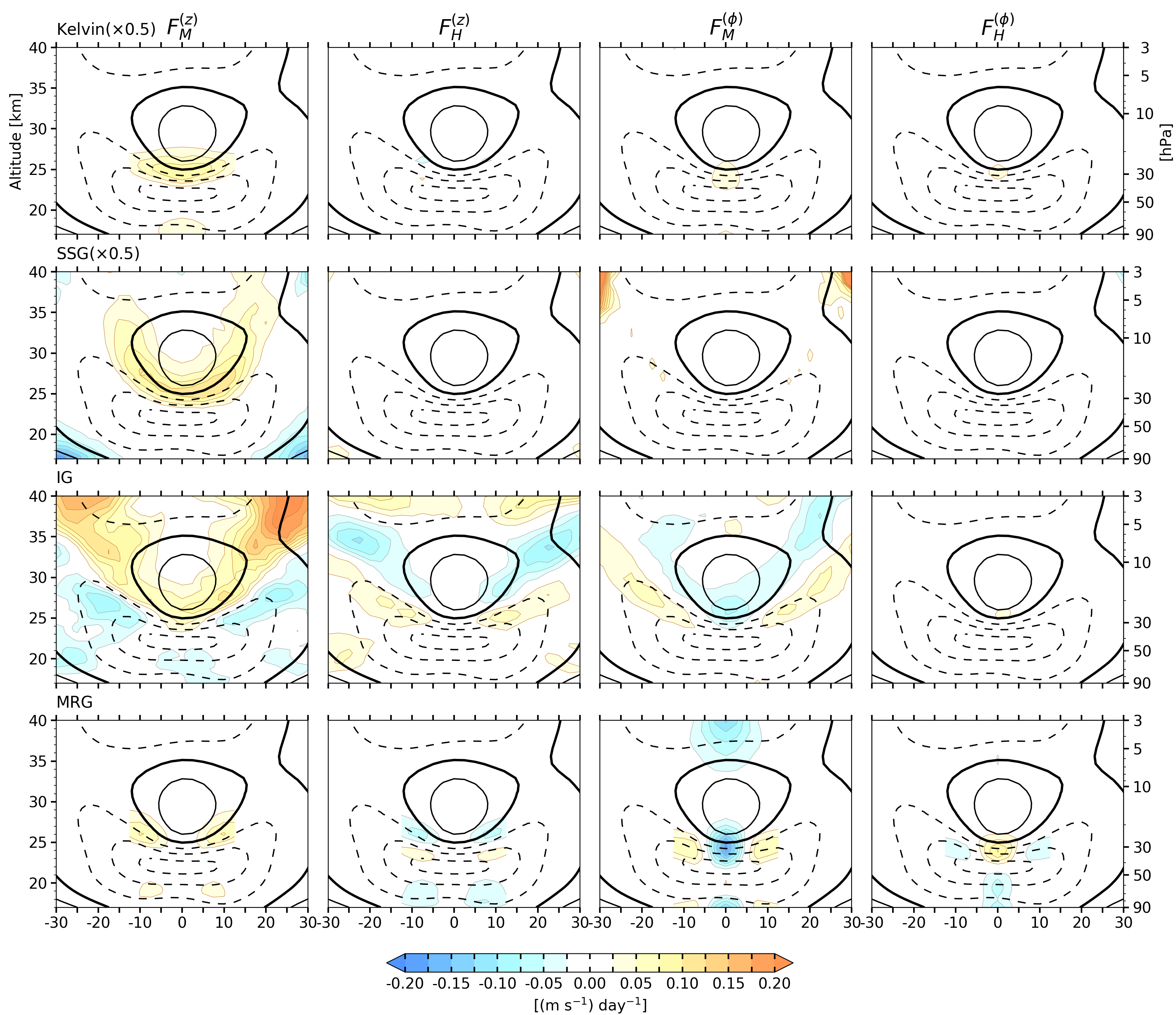}}

\caption{As in Fig. 11, but the forcing of each wave is decomposed into the contributions of each term of the EP flux in the composite for the descending westerly shear zone.} \label{fig12}
\end{figure*}

\begin{figure*}[t]
\centerline{\includegraphics[width=38pc]{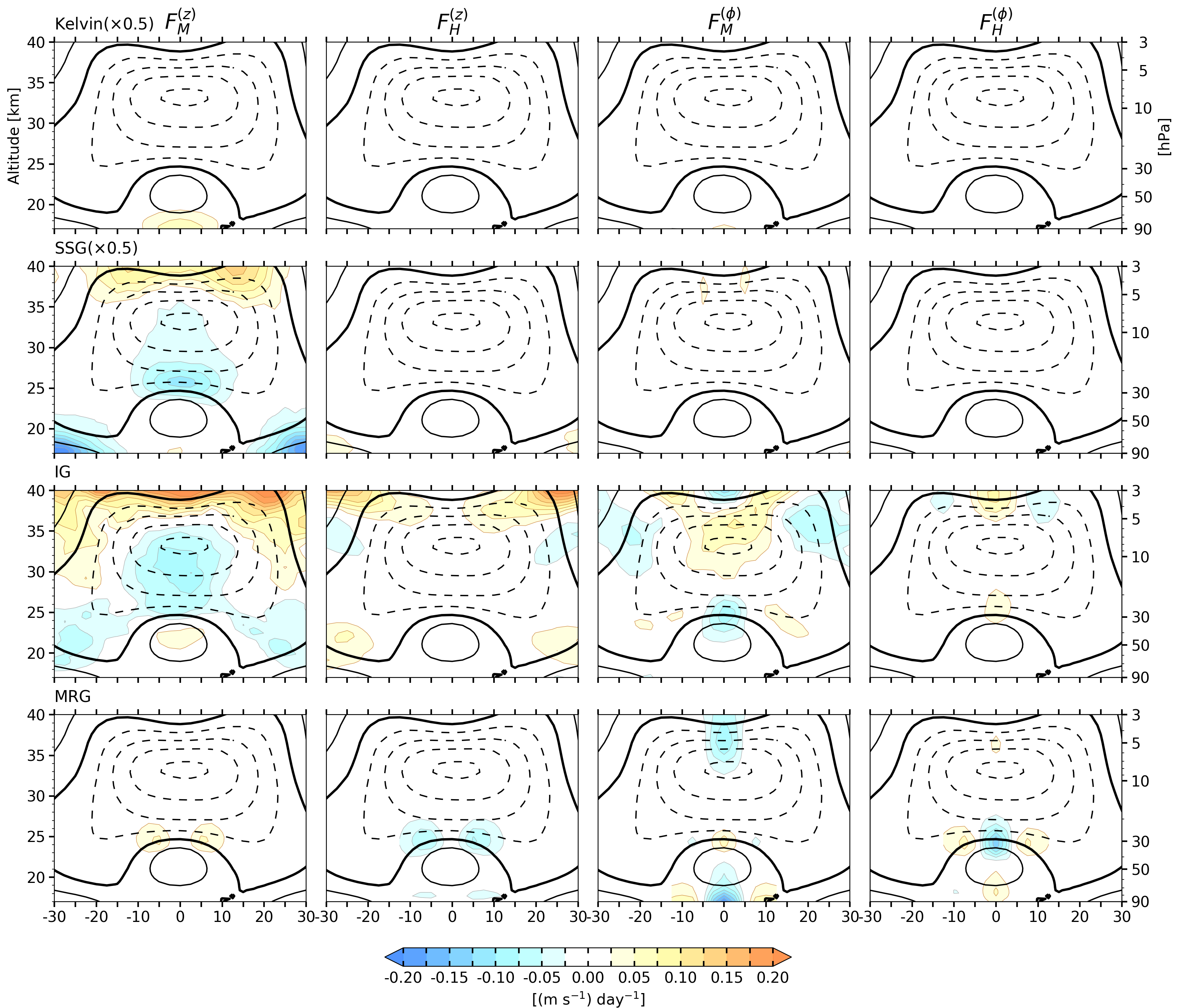}}

\caption{As in Fig. 12, but for composite for the descending easterly shear zone.} \label{fig13}
\end{figure*}

The accelerations due to SSG waves range as high as $\pm$0.3 m s$^{-1}$ d$^{-1}$ ($\pm$0.4 m s$^{-1}$ d$^{-1}$) in strong shear zones around 30 hPa (10 hPa) pressure level (Fig. 10), and are almost entirely a result of the vertical flux of zonal momentum (Figs. 12 and 13). These values are generally consistent with the results from (Kawatani et al. 2010a,b), who used a high-resolution atmospheric general circulation model (60 km horizontal and 300 m vertical resolution), in which waves with short horizontal wavelengths were modeled explicitly and no gravity wave parameterization was required. They estimated accelerations of about $\pm$0.4 m s$^{-1}$ d$^{-1}$ due to the gravity waves with $k > 11$ (wavelengths smaller than $\sim$3600 km) averaged over 10\textdegree  N/S, at 30 hPa, based on which they concluded that the gravity waves with zonal wavelengths $\leq 1000$ km are the main contributors to the descent of the easterly shear zones in the simulated QBO. We refer the reader to Figs. 11 and 12 of Part I of this study for a comparison between the contributions of SSG waves and planetary-scale waves to the forcing of the QBO in ERA5.

The IG waves contribute a small eastward forcing below 10 hPa during the descent of the westerly shear zones, while they play a more significant role in the descent of the easterly shear zones (Figs. 10 and 11). The propagation pattern of the IG waves is different from that in SSG waves (Fig. 11). During the descent of the westerly shear zones, the easterly background wind in the lower stratosphere increases the intrinsic frequencies of the eastward propagating IG waves, thus forcing the IG waves to propagate more vertically. As the IG waves approach the westerly shear zone, their intrinsic frequencies become smaller and the waves tend to propagate more meridionally, depositing most of their eastward momentum off the equator as they dissipate. In contrast, the IG waves do not show substantial vertical propagation during the descent of the easterly shear zones, consistent with results of Kim and Chun (2015b). These IG waves propagate equatorward, converge within the westerly jet and deposit most of their westward momentum when they dissipate in the easterly shear zone above it.

Decomposing the EP flux also reveals another contributing factor to the weak eastward forcing by the IG waves (Figs. 12 and 13): during the descent of westerly shear zones, the accelerations due to $F_M^{(\phi)}$ and $F_H^{(z)}$ oppose that of the $F_M^{(z)}$, leading to an overall weak eastward acceleration along the westerly shear zone (Fig. 11). In contrast, in descending easterly shear zones, $F_M^{(z)}$ induces a westward acceleration, strongest within the easterly regimes, which is reinforced by $F_M^{(\phi)}$ in the easterly shear zone, close to the equator (Fig. 13). The combined forcing results in a relatively strong westward acceleration, particularly close to the equator (Fig. 11).

The MRG waves induce only a weak westward acceleration near the equator in both westerly and easterly shear zones with comparable magnitude (Figs. 10 and 11), suggestive of in situ generation owing to the instability of the QBO westerly jet as discussed (e.g., Maury and Lott 2014; Kim and Chun 2015b; Garcia and Richter 2019). In agreement, it was also shown by Giorgetta et al. (2002, 2006) and Kawatani et al. (2010a,b) that MRG waves play a minor role in the descent of the easterly shear zones in general circulation models.

During the descent of the westerly shear zones, the MRG waves induce eastward accelerations at around 10\textdegree  N/S and westward accelerations near the equator. In descending easterly shear zones, the MRG waves converge near the equator, while the vertical component of the EP flux is quite small, suggesting the stratospheric generation of the MRG waves rather than upward dispersion from the troposphere, consistent with results of Kim and Chun (2015b) based on numerical experiments with the Hadley Centre Global Environment Model version 2 (HadGEM2).

It is evident from Figs. 12 and 13 that MRG westward forcing is due to $F_M^{(\phi)}$ in the westerly shear zone, while it is mostly resulted from $F_H^{(\phi)}$ in the easterly shear zone. Figure 13 shows that the ratio of $F_M^{(z)}$ to $F_H^{(z)}$ in MRG waves is approximately 1 to -2  which is consistent with the theoretical estimation (Andrews et al. 1987).

Figure 10 shows that the Rossby wave forcing is strong in the upper stratosphere, where it tends to occur in phase with easterly accelerations in the semiannual oscillation, but not always. The corresponding meridional cross section shown in Fig. 11 shows extratropical Rossby waves dispersing into the tropical stratosphere and dissipating, inducing a westward acceleration. We have confirmed that the dispersion is almost exclusively from the winter hemisphere, where the waves are able to disperse upward and equatorward from their tropospheric sources through the westerly waveguide (not shown). It is evident from Fig. 11 that the latitude to which the Rossby waves are able to penetrate is modulated by the QBO. They penetrate deeper into westerly wind regimes, and dissipate along the outer edges of easterly regimes, broadening them by extracting westerly momentum. Therefore, although they are not essential to the mechanisms that give rise to the QBO, Rossby waves evidently influence the QBO, as suggested by Dickinson (1968). On the basis of results of experiments with a simple numerical model Dunkerton (1983) argued that Rossby waves can indeed provide part of the forcing of the easterly phase of QBO.

A significant westward forcing is discernible in $F_M^{(\phi)}$ panel of Fig. 8 and Rossby waves panel of Fig. 10 just before the disruption of the QBO in early 2016 when an easterly jet unexpectedly formed in the lower stratosphere at $\sim$40 hPa during an westerly phase. Newman et al. (2016), Osprey et al. (2016), and Coy et al. (2017) have suggested that this disruption was due to the propagation of Rossby waves from the Northern Hemisphere (NH) midlatitudes into the tropical region. Barton and McCormack (2017) argued that NH subtropical easterly jet was anomalously weak during this event due to the timing of the QBO relative to the annual cycle and an exceptionally strong El Nino, which allowed Rossby waves from the NH midlatitudes to penetrate all the way to the equator near the 40 hPa level. Lin et al. (2019) argued that MRG waves also made an appreciable contribution to the disruption.

A second, similar QBO disruption, which began in September 2019 (not shown) has also been attributed to the westward forcing by dissipating Rossby waves from the winter hemisphere, but this time from the Southern Hemisphere (Anstey et al. 2020). These two disruptions suggest that extratropical Rossby waves may be responsible for at least some of the irregularities in the QBO cycle.

\section{Discussion and conclusion}

In this paper we have documented QBO-related wave-mean flow interactions in ERA5. ERA5, which has high spatial and particularly vertical resolution, makes it possible to resolve a broader spectrum of atmospheric waves and to better capture their interaction with the mean flow.

Filtering the resolved waves based on frequency and zonal wavelengths, we have documented the contribution of various types of waves to the forcing of the QBO. The descent of the westerly shear zones is mainly attributable to the Kelvin waves, with a significant contribution of SSG waves, both of which contribute mainly by way of the vertical flux of zonal momentum ($F_M^{(z)}$). That the eastward acceleration by the eastward propagating IG waves is weak below 10 hPa (Fig. 10), is attributable to the fact that their contributions by $F_M^{(z)}$ tend to cancel by way of the vertical flux of heat ($F_H^{(z)}$) and by way of the meridional flux of zonal momentum ($F_M^{(\phi)}$). The westward acceleration is mainly due to the SSG waves, with smaller contributions from IG and MRG waves (Figs. 10 and 11). These results are in agreement with those of other studies based on general circulation models (e.g., Giorgetta et al. 2006; Evan et al. 2012; Krismer et al. 2013) and also with observational studies (e.g., Ern and Preusse 2009a; Alexander and Ortland 2010; Ern et al. 2014), which indicate that small-scale gravity waves are just as important in forcing the descent of QBO westerly wind regimes as Kelvin waves, and that they dominate the wave forcing of easterly regimes.

Somewhat different results were obtained by Garcia and Richter (2019) (GR) based on experiments with the Whole Atmosphere Community Climate Model (WACCM). Their results suggest that the westward acceleration near the equator is due mainly to westward propagating MRG waves with zonal wavenumbers in the range $k = 4-12$ which are generated in situ due to barotropic instability of the QBO westerly jet. The MRG waves in WACCM produce a pattern of EP flux divergence that includes strong westward acceleration close to the equator and eastward acceleration farther from the equator (their Fig. 9), similar to our MRG acceleration pattern in the westerly shear zones (our Fig. 11). As a possible explanation of the difference between their results and those of Kawatani et al. (2010a,b), they suggested that averaging accelerations over more than 5\textdegree  of latitude might have obscured the forcing due to the MRG waves in the previous studies. However, this is not the case in the present study, because, like them, we meridionally average over only 5\textdegree  of latitude, yet we still find that the westward acceleration due to the MRG waves is negligible, even though we include the eastward portion of the $n=0$ continuum in our definition of the MRG space-time domain.

A possible explanation of the differences between GR's results based on WACCM and the results based on simulations using models with higher spatial resolution is the interaction between resolved and parameterized waves. The resolved waves act to prevent instability in the flow driven by the parameterized gravity wave driving (Cohen et al. 2013). GR's results suggest that the MRG waves in WACCM are indeed generated in situ to prevent/reduce the excess of parameterized gravity wave drag in the WACCM that causes the QBO westerly jet to become unstable (their Fig. 7b). If the drag by the unresolved waves is not really as strong as its representation in WACCM, then the MRG waves found by GR might not exist in nature. The forcing by the resolved waves in our study should be less subject to this artificial compensation problem. Even if the gravity wave parameterization in ERA5 is not perfect, the analyzed dynamical fields, including the resolved waves, should be reasonably realistic because they are based on extensive observational data that have been assimilated into the model using state-of-the-art data assimilation techniques  (Sato and Hirano 2019). Furthermore, the parameterized gravity wave drag in ERA5 appears to act to reduce, rather than create instabilities in the zonally symmetric flow in the tropical stratosphere, as discussed in Part I.

\acknowledgments
This research is supported by the NASA grant 80NSSC18K1031 and NSF grant AGS-1821437. The authors declare no conflicts of interest.

\datastatement
ERA5 data were downloaded from ECMWF's MARS archive. Aura MLS observations were obtained from NASA Goddard Earth Sciences (GES) Data and Information Services Center (DISC). CALIPSO data were downloaded using NASA Search and Subsetting Web Application (https://subset.larc.nasa.gov/calipso/).

\bigskip

 \textbf{References}

\bigskip

Alexander, M. J., and D. A. Ortland, 2010: Equatorial waves in High Resolution Dynamics Limb Sounder (HIRDLS) data. Journal of Geophysical Research: Atmospheres, 115.

Andrews, D. G., J. D. Mahlman, and R. W. Sinclair, 1983: Eliassen-Palm Diagnostics of Wave-Mean Flow Interaction in the GFDL SKYHI General Circulation Model. J. Atmos. Sci., 40, 2768-2784, https://doi.org/10.1175/1520-0469(1983)040<2768:ETWATM>2.0.CO;2.

Andrews, D. G., C. B. Leovy, and J. R. Holton, 1987: Middle atmosphere dynamics. Academic press,.

Anstey, J. A., T. P. Banyard, N. Butchart, L. Coy, P. A. Newman, S. Osprey, and C. Wright, 2020: Quasi-biennial oscillation disrupted by abnormal Southern Hemisphere stratosphere. Earth and Space Science Open Archive,. http://www.essoar.org/doi/10.1002/essoar.10503358.1 (Accessed July 8, 2020).

Baldwin, M. P., and Coauthors, 2001: The quasi-biennial oscillation. Reviews of Geophysics, 39, 179-229.

Barton, C. A., and J. P. McCormack, 2017: Origin of the 2016 QBO disruption and its relationship to extreme El Niño events. Geophysical Research Letters, 44, 11-150.

Cohen, N. Y., Gerber, and E. P. Oliver Bühler, 2013: Compensation between resolved and unresolved wave driving in the stratosphere: Implications for downward control. Journal of the atmospheric sciences, 70, 3780-3798.

Coy, L., P. A. Newman, S. Pawson, and L. R. Lait, 2017: Dynamics of the disrupted 2015/16 quasi-biennial oscillation. Journal of climate, 30, 5661-5674.

Dias, J., and G. N. Kiladis, 2016: The relationship between equatorial mixed Rossby-gravity and eastward inertio-gravity waves. Part II. Journal of the Atmospheric Sciences, 73, 2147-2163.

Dickinson, R. E., 1968: Planetary Rossby waves propagating vertically through weak westerly wind wave guides. Journal of the Atmospheric Sciences, 25, 984-1002.

Dunkerton, T. J., 1983: Laterally-propagating Rossby waves in the easterly acceleration phase of the quasi-biennial oscillation. Atmosphere-Ocean, 21, 55-68.

——, 1997: The role of gravity waves in the quasi-biennial oscillation. Journal of Geophysical Research: Atmospheres, 102, 26053-26076.

Ern, M., and P. Preusse, 2009a: Quantification of the contribution of equatorial Kelvin waves to the QBO wind reversal in the stratosphere. Geophysical research letters, 36.

——, and ——, 2009b: Wave fluxes of equatorial Kelvin waves and QBO zonal wind forcing derived from SABER and ECMWF temperature space-time spectra. Atmos. Chem. Phys, 9, 3957-3986.

——, and Coauthors, 2014: Interaction of gravity waves with the QBO: A satellite perspective. Journal of Geophysical Research: Atmospheres, 119, 2329-2355.

Evan, S., M. J. Alexander, and J. Dudhia, 2012: WRF simulations of convectively generated gravity waves in opposite QBO phases. Journal of Geophysical Research: Atmospheres, 117.

Garcia, R. R., and J. H. Richter, 2019: On the momentum budget of the quasi-biennial oscillation in the whole atmosphere community climate model. Journal of the Atmospheric Sciences, 76, 69-87.

Giorgetta, M. A., E. Manzini, and E. Roeckner, 2002: Forcing of the quasi-biennial oscillation from a broad spectrum of atmospheric waves. Geophysical Research Letters, 29, 86-1.

Giorgetta, M. A., E. Manzini, E. Roeckner, M. Esch, and L. Bengtsson, 2006: Climatology and forcing of the quasi-biennial oscillation in the MAECHAM5 model. Journal of Climate, 19, 3882-3901.

Hersbach, H., and Coauthors, 2020: The ERA5 global reanalysis. Quarterly Journal of the Royal Meteorological Society, 146, 1999-2049, https://doi.org/10.1002/qj.3803.

Holt, L. A., and Coauthors, 2020: An evaluation of tropical waves and wave forcing of the QBO in the QBOi models. Quarterly Journal of the Royal Meteorological Society, n/a, https://doi.org/10.1002/qj.3827.

Holton, J. R., and R. S. Lindzen, 1972: An updated theory for the quasi-biennial cycle of the tropical stratosphere. Journal of the Atmospheric Sciences, 29, 1076-1080.

Kawatani, Y., S. Watanabe, K. Sato, T. J. Dunkerton, S. Miyahara, and M. Takahashi, 2010a: The roles of equatorial trapped waves and internal inertia-gravity waves in driving the quasi-biennial oscillation. Part II: Three-dimensional distribution of wave forcing. Journal of the atmospheric sciences, 67, 981-997.

——, ——, ——, ——, ——, and ——, 2010b: The roles of equatorial trapped waves and internal inertia-gravity waves in driving the quasi-biennial oscillation. Part I: Zonal mean wave forcing. Journal of the atmospheric sciences, 67, 963-980.

Kiladis, G. N., J. Dias, and M. Gehne, 2016: The relationship between equatorial mixed Rossby-gravity and eastward inertio-gravity waves. Part I. Journal of the Atmospheric Sciences, 73, 2123-2145.

Kim, Y.-H., and H.-Y. Chun, 2015a: Momentum forcing of the quasi-biennial oscillation by equatorial waves in recent reanalyses. Atmospheric Chemistry \& Physics, 15.

——, and ——, 2015b: Contributions of equatorial wave modes and parameterized gravity waves to the tropical QBO in HadGEM2. Journal of Geophysical Research: Atmospheres, 120, 1065-1090.

Kim, Y.-H., and Coauthors, 2019: Comparison of equatorial wave activity in the tropical tropopause layer and stratosphere represented in reanalyses. Atmospheric Chemistry and Physics, 19, 10027-10050, https://doi.org/10.5194/acp-19-10027-2019.

Krismer, T. R., M. A. Giorgetta, and M. Esch, 2013: Seasonal aspects of the quasi-biennial oscillation in the Max Planck Institute Earth System Model and ERA-40. Journal of Advances in Modeling Earth Systems, 5, 406-421.

Lin, P., I. Held, and Y. Ming, 2019: The early development of the 2015/16 quasi-biennial oscillation disruption. Journal of the Atmospheric Sciences, 76, 821-836.

Lindzen, R. S., and J. R. Holton, 1968: A theory of the quasi-biennial oscillation. Journal of the Atmospheric Sciences, 25, 1095-1107.

Lott, F., and Coauthors, 2014: Kelvin and Rossby-gravity wave packets in the lower stratosphere of some high-top CMIP5 models. Journal of Geophysical Research: Atmospheres, 119, 2156-2173, https://doi.org/10.1002/2013JD020797.

Matsuno, T., 1966: Quasi-Geostrophic Motions in the Equatorial Area. Journal of the Meteorological Society of Japan. Ser. II, 44, 25-43, https://doi.org/10.2151/jmsj1965.44.1-25.

Maury, P., and F. Lott, 2014: On the presence of equatorial waves in the lower stratosphere of a general circulation model. Atmospheric Chemistry and Physics, 14, 1869-1880, https://doi.org/10.5194/acp-14-1869-2014.

Newman, P. A., L. Coy, S. Pawson, and L. R. Lait, 2016: The anomalous change in the QBO in 2015-2016. Geophysical Research Letters, 43, 8791-8797.

Osprey, S. M., N. Butchart, J. R. Knight, A. A. Scaife, K. Hamilton, J. A. Anstey, V. Schenzinger, and C. Zhang, 2016: An unexpected disruption of the atmospheric quasi-biennial oscillation. Science, 353, 1424-1427.

Richter, J. H., J. A. Anstey, N. Butchart, Y. Kawatani, G. A. Meehl, S. Osprey, and I. R. Simpson, 2020: Progress in Simulating the Quasi-Biennial Oscillation in CMIP Models. Journal of Geophysical Research: Atmospheres, 125, e2019JD032362, https://doi.org/10.1029/2019JD032362.

Salby, M. L., and R. R. Garcia, 1987: Transient response to localized episodic heating in the tropics. Part I: Excitation and short-time near-field behavior. Journal of the atmospheric sciences, 44, 458-498.

Sato, K., and S. Hirano, 2019: The climatology of the Brewer-Dobson circulation and the contribution of gravity waves. Atmos. Chem. Phys, 19, 4517-4539.

Schenzinger, V., S. Osprey, L. Gray, and N. Butchart, 2017: Defining metrics of the Quasi-Biennial Oscillation in global climate models. Geoscientific Model Development, 10.

Stephan, C., and M. J. Alexander, 2015: Realistic simulations of atmospheric gravity waves over the continental U.S. using precipitation radar data. Journal of Advances in Modeling Earth Systems, 7, 823-835, https://doi.org/10.1002/2014MS000396.

Vincent, R. A., and M. J. Alexander, 2020: Observational studies of short vertical wavelength gravity waves and interaction with QBO winds. Earth and Space Science Open Archive, https://doi.org/10.1002/essoar.10502563.1.

Wheeler, M., and G. N. Kiladis, 1999: Convectively Coupled Equatorial Waves: Analysis of Clouds and Temperature in the Wavenumber-Frequency Domain. J. Atmos. Sci., 56, 374-399, https://doi.org/10.1175/1520-0469(1999)056<0374:CCEWAO>2.0.CO;2.

\end{document}